\documentclass[prb,twocolumn,showpacs,a4paper]{revtex4}

\usepackage{graphicx}
\usepackage{dcolumn}
\usepackage{bm}

\begin{document}

\preprint{APS/123-QED}

\title{Size effects on the quenching to the normal state  in $\rm{YBa_2Cu_3O_{7-\delta}}$ thin film superconductors.}

\author{Mauricio Ruibal}
 \email{fmruibal@usc.es}
 \homepage{http://aiff.usc.es/~mclbt}


\author{Gonzalo Ferro}

\author{Manuel R. Osorio}
 
\author{Jes\'us Maza}

\author{Jos\'e A. Veira}

\author{F\'elix Vidal}
 \affiliation{LBTS, Facultade de F\'isica, Universidade de Santiago de Compostela, E15782, Spain.}


\date{\today}

\begin{abstract}

To probe the quenching mechanisms under high current densities,  current-voltage curves have been measured in $\rm{YBa_2 Cu_3 O_{7-\delta}}$ thin film microbridges with widths lower than the thermal diffusion length. This condition was obtained by using microbridge widths under 100 $\mu$m  and stepped ramps of one millisecond step duration. Whereas the flux-flow resistivity is found to be microbridge-width independent, strong width dependence of the  quenching current density is observed. These results provide a direct experimental demonstration that for high current densities varying in the millisecond range the transition to a highly dissipative state  is due to self heating driven by ``conventional'' (non-singular) flux flow effects.

\end{abstract}
\maketitle

   It is now well established in low  and high-$T_c$ superconductors that high current densities may induce an abrupt transition from the superconducting state to a highly dissipative regime (up to the normal state) \cite{Tinkham, Gurevich1987,Kunchur2004,Altshuler2004}. This transition manifests itself in the current-voltage (C-V) curves as an abrupt jump in voltage at some current density, denoted $J^*$, several times the critical current density, $J_c$, at which dissipation first appears. This phenomenon, often called quenching, is still not well understood, in spite of its considerable importance from both the fundamental and the applied point of view \cite{Tinkham, Gurevich1987,Kunchur2004,Altshuler2004}. The main mechanisms    proposed  to explain the quenching of the superconductivity by high current densities include nonlinear vortex dynamics \cite {LO1976,BS1992,Doettinger1994,Klein1985,Bernstein2005,Altshuler2004,Villard2005,Samoilov1995,Ruck1997}, Cooper pair breaking \cite{Kunchur2004,Sabouret2002,Reymond2002}, hot-spots \cite{Xiao1998, Maneval2001} and self heating \cite{Kiss1999,Lehner2002,Jose2003,Gurevich1987}. The measurements performed in the last years by various groups showed that  these different mechanisms do not exclude each other \cite{Xiao1998,Sabouret2002,Antognazza2001,Teresa2003,Jose2003,Kunchur2002,Kunchur2004,Samoilov1995,Ruck1997}. In fact, these experiments suggest that depending on the properties of a given superconductor material and on the experimental conditions but also on the way  the  current is supplied,  each one of these quenching mechanisms may become the most relevant one  or they may cooperate  to reach the high dissipative regime.

\begin{figure}[hb]
	\centering
		\includegraphics[width=0.35\textwidth]{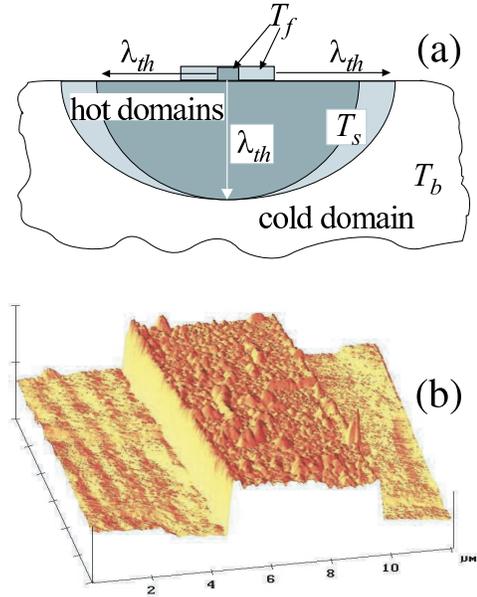}
	\caption{(a) Schematic  temperature distribution  of the film-substrate system (cross-section) for two microbridges with different widths but, in both cases, with widths lower than the thermal diffusion length of their substrate. Under this condition, the narrower microbridge will be much better refrigerated by the substrate than the wider microbridge. (b) AFM image of a 5 $\mu m$ width microbrige.}
	
	\label{fig:diagrama}
\end{figure} 

In this paper, we report an experiment addressed to unambiguously separate self heating from other quenching mechanisms and to probe if self-heating \emph{alone}, associated with conventional (non singular) flux flow, could originate the abrupt transition to the normal state. This experiment is based on a very simple idea schematized in Fig.~\ref{fig:diagrama}(a): for a superconducting thin film with width, $w$, below the thermal diffusion length of both the  superconductor film itself and its substrate, $\lambda_{th}$, the  volume of the substrate appreciably heated will decrease less than the film volume when the film width decreases. Since the thermal exchanges with the substrate are the more relevant ones, independently of the refrigeration atmosphere \cite{Jose2003,Gupta1993}, the constraint $w < \lambda_{th}$ will determine the entire film thermal behaviour, in particular the temperature increase $\Delta T_f$ between the film and its bath. Although  below $J^*$ the width dependence of $\Delta T_f$ will be quite smooth [in our experiments $\Delta T_f (w)$ will change less than 4 \% of the bath temperature, $T_b$], such a width dependence  dramatically affects the self heating and then the corresponding $J^* (w)$. In contrast, not too close to $T_c$ (for $T$ well below $T_c - \Delta T_f$), these $\Delta T_f (w)$ variations are too small to appreciably affect other quenching mechanisms. In addition, the condition $w < \lambda_{th}$ may be easily verified by keeping the characteristic times and lengths of these other quenching mechanisms completely decoupled from $w$ and  the current-density variation times. Therefore,  under these conditions, the presence of appreciable size effects on $J^*$ will provide a fingerprint for the self-heating mechanism.

In our experiments, we use high-quality c-axis oriented $\rm{YBa_2Cu_3O_{7-\delta}}$ (YBCO) thin film microbridges patterned with different widths (between \mbox{5 $\mu$m} and \mbox{100 $\mu$m}) but  the same width-to-length ratio (1/10) and thickness (\mbox{120 nm}). The transition to the normal state is induced under zero applied magnetic field and \mbox{$\mu_0 H= 1$ T}, at temperatures not too far below the superconducting transition (typically,  about twenty degrees below $T_c$), and by using stepped ramps of high current density of total time in the range of tens of milliseconds and steps of one millisecond duration. Under these conditions, the film width remains well below the thermal diffusion length, $\lambda_{th} = 2(Dt)^{1/2}$, of both the YBCO microbridges and their substrate ($\rm SrTiO_3$). Here $D$ is the corresponding thermal diffusivity and $t$ is the current step time. Under low applied magnetic fields, $D$ is of the order of \mbox{0.05 cm$^2$/s} for YBCO, and around \mbox{0.2 cm$^2$/s} for the $\rm SrTiO_3$ substrates. This leads to $\lambda_{th} \approx 150$ $\mu$m and \mbox{250 $\mu$m} for the YBCO microbridges and, respectively, their substrates. Therefore, in this scenario not only the film heating will become uniform but also the characteristic lengths and times for other possible quenching mechanisms are much smaller or bigger than the microbridge width  and the current-variation times in our experiments \cite{longitudes}. In contrast, we will see that the microbridges studied here covered all the interesting $\lambda_{th}/w > 1$ region corresponding to the millisecond range.

Three other aspects of our present experiments deserve to be stressed already here. First, to discriminate the size  effects on $J^* (w)$, it was crucial to obtain high-quality microbridges with different widths below $\lambda_{th}$, but with the same electrical resistivity, $\rho$ (which will guarantee that the different microbridges have similar stoichiometric and structural characteristics), and also with the same flux-flow resistivity, $\rho _f$ (which will guarantee a uniform current distribution in the dissipative regime above $J_c$). Second, although some of the previous experimental studies suggest the relevance of the thermal effects near $J^*$ \cite{Kiss1999,Lehner2002,Jose2003}, up to now it does not exist an experimental demonstration that in some cases self heating \emph{by itself}, associated with non singular flux flow, may generate the quenching. As it was stressed before, the scenario studied here will allow to directly probe, for the very first time, such a thermal quenching mechanism. In addition to its interest for the understanding of the transport properties in superconductors, our present work will provide  one of the very few existing experimental demonstrations of a thermal avalanche  to a highly dissipative state in any material, a phenomena directly related to bistability  and dynamic avalanches in a large variety of systems\cite{Gurevich1987, Altshuler2004}. Finally, the millisecond range studied here corresponds to various of the most promising applications of the cuprate high-$T_c$ superconductors. This is the case, for instance, of the fault-current limiters, the practical fault to be limited having typical characteristic times of the order of the commercial ac currents, i.e., $\sim 10$ ms \cite{Limitador}.

\begin{figure}[t]
	\centering
		\includegraphics[width=0.4\textwidth]{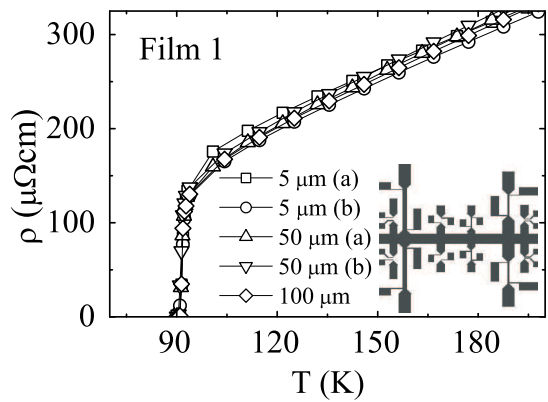}
	\caption{Electrical resistivity versus temperature curves of several microbridges from the film 1. The two different microbridges of same width are noted (a) and (b) and the numbers stand for the films widths. In the inset  the geometrical design of these microbridges is shown. The widths are $w$ =  5, 10, 20, 50 and 100 $\mu$m. }
	\label{fig:resistividad}
\end{figure}

The epitaxial YBCO thin films of  thickness \mbox{$d\approx$ 120 nm} were grown onto $\rm SrTiO_{3}$(100) substrates by pulsed laser ablation and the growth was monitored with a RHEED  system \cite {Rijnders1997}. In our experiments we have used two films (noted 1 and 2). Several microbridges of widths $w$ = 5, 10, 20, 50 and 100 $\mu$m   were patterned on each film by photolithography and etched with  Ar ions into a four-probe configuration as  can be seen in the inset of Fig.~\ref {fig:resistividad}. The C-V curves were obtained by applying stepped ramps of about \mbox{1 ms} of step duration. Other experimental details are similar to those described for other measurements in Ref.~\cite{Jose2003}.

\begin{figure}[t]
	\centering
	\includegraphics[width=0.35\textwidth]{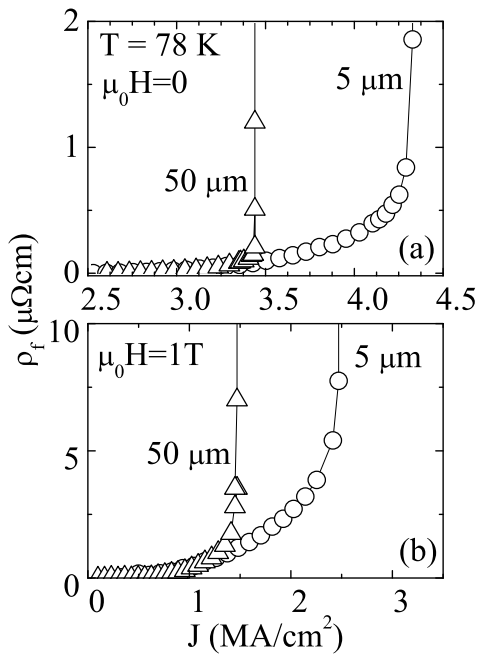}
					\caption{Flux-flow resistivity as a function of the applied current density up to the quenching current for two of the microbridges studied here without applied magnetic field, (a), and with 1~T, (b). These examples, which correspond to two microbridges with widths of \mbox{5 $\mu$m} and \mbox{50 $\mu$m} from film 1, directly show two of the central aspects of this work: i) below $J^*$, the flux-flow resistivity is width independent, ii) $J^*$ increases as the film width decreases.}
\label{fig:fluxflow}
\end{figure}

\begin{figure}[t]
	\centering
		\includegraphics[width=0.35\textwidth]{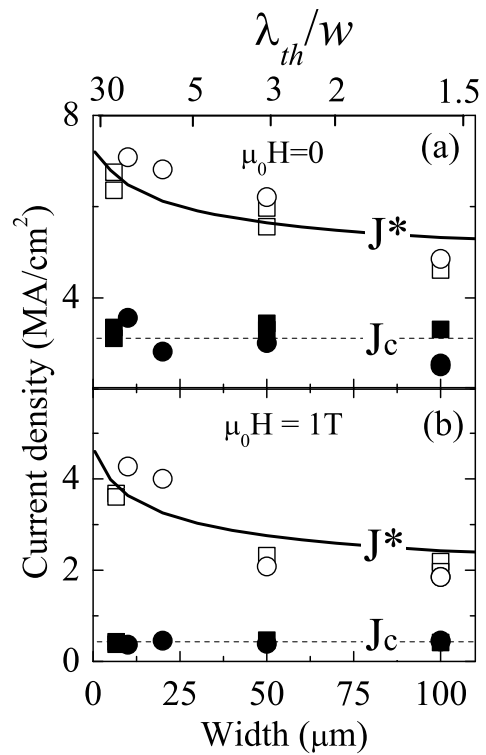}	
			\caption{Microbridge width dependence of the quenching current density $J^*$ (open symbols) in absence of applied magnetic field (a) and under 1~Tesla (b), both at $T/T_c = 0.8$. For comparison, we also show in this figure $J_c$ (defined using the 10 $\mu V/cm$ standard criterion for films) at the same $T/T_c$ (solid symbols). The upper scale shows that our experiments cover all the relevant $\lambda_{th} / w > 1$ region corresponding to the millisecond range.  The solid lines are estimated from the self-heating model described in the text. The dotted lines are just lines for the eyes.} 
			
\label{fig:vsanchos}
\end{figure}

Some examples of the electrical resistivity versus temperature curves, measured under dc conditions are presented in Fig.~\ref{fig:resistividad}. These data clearly show that $T_c$ and $\rho (T)$ are almost width independent. Various examples of the flux-flow resistivity versus current density curves are shown in Fig.~\ref{fig:fluxflow}. The data points correspond to \mbox{$T =$ 78 K}, with $\mu_0 H= 0$ or \mbox{$\mu_0 H= 1$ T}, and they were obtained by dividing the measured electric field of each step by the corresponding current density \cite{Comentario}. These  curves provide already an illustrative example of the strong dependence of the quenching current density, $J^*$, on the microbridge width. In contrast, one may also see that below $J^*$ the flux-flow resistivity, $\rho _f$, for the two microbridges agree with each  other well within the experimental resolution. This last result provides a direct evidence that in the dissipative regime the current flows uniformly through the microbridges, which is one of the starting hypothesis of the approach we will use below to explain the width dependence of $J^*$ in terms of self-heating effects. This independence of $\rho _f(J)$ on $w$ was already observed  by Kunchur and co-workers in YBCO thin films with \mbox{3 $\mu$m $\lesssim w \lesssim$ 16 $\mu$m} under different applied magnetic field amplitudes and using current pulses in the microsecond range \cite{Kunchur2000}. It may be easily understood by taking into account that the transversal magnetic field penetration length of YBCO is much larger than the film's thickness and also that in the dissipative regime the current distribution is dominated by the minimum entropy production principle. In fact, we have also performed detailed dc measurements in the low-dissipation regime in our different microbridges that confirms that even $J_c$ is almost width independent. Some examples of these $J_c (w)$ and also of the corresponding $J^* (w)$ data are shown in Fig.~\ref{fig:vsanchos}.  These examples were obtained at $T/T_c = 0.8$, which corresponds to \mbox{$T =$ 73 K} and \mbox{$T =$ 71 K} for, respectively, the films 1 (squares) and  2 (circles).  The data points scattering of both $J_c$ and $J^*$ corresponds well to the experimental uncertainties. These last are associated with the samples geometry but also with differences in the microbridges surface roughness [see Fig. \ref{fig:diagrama}(b)], which could appreciably affect the vortex pinning. To take into account the small differences between the two films, the $J_c$ and $J^*$ data for the microbridges from film 2 have been normalized by their resistivity over the resistivity of the film 1 microbridges. Such a renormalization may increase the differences between the data points corresponding to microbridges of the two different films. 

The absolute value of $J^*$ for the \mbox{10 $\mu$m} width microbridge  showed in Fig.~\ref{fig:vsanchos} agrees qualitatively with the  existing data obtained under similar scenarios in YBCO films \cite{Xiao1998,Jose2003}. The new central result of our present work is the microbridge width dependence of $J^*$: as showed in the two examples presented in Fig.~\ref{fig:vsanchos},  $J^* (w)$ appreciably increases when the width of the microbridges decreases well below the substrate thermal diffusion length. The relative $J^* (w)$ changes are at about 30~\% and 80~\% under, respectively, zero applied field and 1~T. This behaviour may be qualitatively understood  in terms of self heating by just taking into account that the narrower microbridges are better refrigerated by their substrate. These results provide then a first qualitative but direct indication that under the studied conditions the quenching mechanism is self heating. 

To analyze at a quantitative level the $J^* (w)$ behavior observed in Fig.~\ref{fig:vsanchos}, we will use a simple two-dimensional self-heating model similar to the one proposed in Ref.~\cite{Jose2003}. As noted before, in this model it is assumed a uniform current distribution through the microbridge volume and the diffusion into the substrate of the heat dissipated by this current is calculated using a simple geometry, discretizing the continuous temperature field into two constant-temperature domains (see Fig.~\ref{fig:diagrama}(a)). A hot one, with temperature $T_s(t)$, its volume being time dependent, and a cold one (the remaining part of the substrate) at the bath temperature, $T_b$. The iterative process for calculating the self heating has three steps. First, the heat dissipated is estimated by using a power relationship for the electric field $E$ versus  the current density: $E = E_0 (J/J_c - 1)^n$ in the case of zero applied magnetic field, and $E= E_c (J/J_c )^s$ under magnetic field, where $E_0$, $J_c$ and $s$ are temperature-dependent and $n$ and $E_c$ are temperature-independent parameters. These parameters are obtained from the measured C-V curves at different temperatures in the low-current (``heatless'') range. Then, the temperature increase of the film $\Delta T_f$ at time $t$ is estimated from the relationship $\Delta T_f = \alpha \Delta T_s (t)$, where $\Delta T_s$ is the temperature increase of the hot domain and $\alpha (t)=1+1.87(1-e^{-t/t_c})$,  $t_c \approx  10$ ns being the bolometric time\cite{tiempo,Jose2003, Gupta1993}. Just before the quenching, in our experiments this leads to $\Delta T_f \sim 2$K in absence of an applied field and to $\Delta T_f \sim 4$K when $\mu_0 H= 1$T. Let us stress that this relatively small temperature increase suffices to trigger the thermal avalanche but does not appreciably affect other possible quenching mechanisms as vortex avalanches. In step three, the power dissipated by the current density $J$ is calculated by using the corresponding value of $E$ at the new temperature. This power value is used in the next iteration. 

As showed in Fig.~\ref{fig:vsanchos}, the $J^* (w)$ curves calculated as indicated before are in quite good agreement with the experimental data, in spite of the crudeness of our 2D self-heating model and that for each applied field we have used the same parameter values for all samples. It is remarkable that such an agreement is found not only for the $J^*$ width-dependence but also for its \textit{absolute} amplitude, and that with no free parameters. Two key ideas behind the self-heating calculations performed here have to be emphasized. First, smooth or non-singular background C-V curves, extrapolated from the low $J$ (i. e., ``heatless'') regime, are used for the calculation of the injected power during the current pulses.  Secondly, the non-linearity in temperature of the C-V curves is taken into account by a thermal feedback along the iterative calculation leading to $J^*$. So, our approach is a purely thermal model.

In conclusion, although to observe strong size effects on $J^*$ it was necessary to use narrow microbridges, with $w < \lambda_{th}$, our results lead to a much more general conclusion, applicable also to larger high-$T_c$ cuprate films: For high current densities varying in the \textit{millisecond range}, the \textit{intrinsic} quenching mechanism is self heating driven by conventional (non-singular) flux-flow effects. This conclusion does not exclude the relevance of other quenching mechanisms  for much shorter characteristic current times \cite{tiempo} and, indeed, with higher $J^*$ values \cite{Klein1985,Doettinger1994,Sabouret2002,Kunchur2004,Xiao1998,Maneval2001,Kunchur2002,LO1976,BS1992,Bernstein2005,Kunchur2000,Villard2005}. In addition to its interest for the understanding of the transport properties of superconductors under high current densities,  these results provide one of the very few existing experimental demonstrations of thermal-driven avalanche in any material\cite{Gurevich1987,Kunchur2004,Altshuler2004}. 


This work has been supported by the CICYT, Spain, under grant no.~MAT2004-04364, Xunta de Galicia under grant PGIDIT04TMT206002PR and by Uni\'on Fenosa under contract 0666-2002.
The films used in this work were grown by M. R. during his stay at the MESA+ research institute (University of Twente, The Netherlands) and thanks are due to  G. Rijnders for his kind help and hospitality.


\begin{thebibliography}{19}

	\bibitem{Tinkham} See, e.g., M. Tinkham, \textit{Non-equilibrium superconductivity, Phonons and Kapitzka boundaries}, edited by K. E. Gray (Plenum, New York, 1981), p. 231.
	
	\bibitem{Gurevich1987} A. VI. Gurevich and R. G. Mints,  Rev. Mod. Phys. {\bf 59}, 941 (1987) and references therein.
	
	\bibitem{Kunchur2004} See, e. g., M. N. Kunchur,  J. of Phys: Cond. Matt. {\bf 16}, R1183 (2004) and references therein.
	
	\bibitem{Altshuler2004} E. Altshuler and T. H. Johansen, Rev. Mod. Phys. {\bf 76}, 471 (2004) and references therein.
	
	\bibitem{Klein1985} W. Klein, R. P. Huebener, S. Gauss, and J. Parisi,  J. Low. Temp. Phys. {\bf 61}, 413 (1985).
			
	\bibitem{Doettinger1994} S. G. Doettinger  \textit{et al.},  Phys. Rev. Lett. {\bf 73}, 1691(1994).
	
	
		\bibitem{LO1976} A. I. Larkin and Y. N. Ovchinnikov,   Sov. Phys. JETP {\bf 41}, 960 (1976).
	
	\bibitem{BS1992} A. I. Bezuglyj  and V A Shklovskij,    Physica C {\bf 202}, 234 (1992).
	
	\bibitem{Villard2005} C. Peroz and C. Villard, Phys. Rev. B {\bf 72}, 014515 (2005).

  \bibitem{Samoilov1995} A. V. Samoilov  \textit{et al.}, Phys. Rev. Lett. {\bf 75}, 4118 (1995).
  
	\bibitem{Ruck1997} B. J. Ruck  \textit{et al.}, Phys. Rev. Lett. {\bf 78}, 3378 (1997).
	
	\bibitem{Bernstein2005} P. Bernstein \textit{et al.},  J. Phys.: Conf. Ser., in press.
	
		
	\bibitem{Sabouret2002} G. Sabouret, C. Williams, and R. Sobolewski,   Phys. Rev. B {\bf 66}, 132501 (2002).
	
	\bibitem{Reymond2002} S. Reymond \textit{et al.}, Phys. Rev. B {\bf 66}, 014522 (2002).
		
	\bibitem{Xiao1998} Z. L. Xiao, E. Y. Andrei, and P. Ziemann,  Phys. Rev. B {\bf 58}, 11185 (1998) and references therein.
	
		\bibitem{Maneval2001} J-P. Maneval \textit{et al.},  J. Supercond. {\bf 14}, 347 (2001); S. Michotte \textit{et al.}, Appl. Phys. Lett. {\bf 85}, 3175 (2004).
	
		
	
	\bibitem{Kiss1999} T. Kiss \textit{et al.},  IEEE Trans. on Appl. Supercon. {\bf 9}, 1073 (1999).
	
	\bibitem{Lehner2002} A. Lehner \textit{et al.},  Physica C {\bf 372-276}, 1619 (2002).
	
	\bibitem{Jose2003} J. Vi\~na \textit{et al.},   Phys. Rev. B {\bf 68}, 224506 (2003).
	
		\bibitem{Kunchur2002} M. N. Kunchur,   Phys. Rev. Lett. {\bf 89}, 137005 (2002).
	
	\bibitem{Antognazza2001} M. Decroux \textit{et al.}, IEEE Trans. on Appl. Supercon. {\bf 11}, 2046 (2001).
	
	\bibitem{Teresa2003} M. T. Gonz\'alez \textit{et al.},  Phys. Rev. B {\bf 68}, 054514 (2003).
	
		\bibitem{Gupta1993} S. K. Gupta \textit{et al.},  Physica C {\bf 206}, 335 (1993).
	
		\bibitem{longitudes} This is the case, specifically, of the two central characteristics lengths for the superconductivity in YBCO: the superconducting coherence length (which take values very small in all directions, between \mbox{0.1 nm} to \mbox{2 nm}) and the magnetic-field penetration length (having values around \mbox{600 nm} in the transversal direction, which are to be compared with \mbox{$d =$ 120 nm} for our films). The film widths are  also much larger than the YBCO phonon mean free path. This guarantees the absence of appreciable width effects associated with  boundary scattering. See, e.g., M. I. Flik and C. L. Tien, ASME J. Heat Transfer {\bf 112}, 872 (1990).
	
	\bibitem{Limitador} See, e.g., T. Verhaege and Y. Laumond,  Handbook of Applied Superconductivity, vol.2 (Institute of Physics Publishing, Bristol, 1998), p. 1691.
	
	
\bibitem{Rijnders1997} G. J. H. M. Rijnders \textit{et al.},  Appl. Phys. Lett. {\bf 70}, 1888 (1997).



	\bibitem{Comentario} In the measurements performed under zero applied field, the magnetic vortices are due to the self-field. See, e.g., M. T. Gonz\'alez, S. R. Curr\'as, J. Maza, and F. Vidal, Phys. Rev. B. {\bf 63}, 224511 (2001), and references therein.
		
	\bibitem{Kunchur2000} M. N. Kunchur, B. I. Ivlev, D. K. Christen, and J. M. Phillips, Phys. Rev. Lett. {\bf 84}, 5204 (2000).
	
	
	
	
\bibitem{tiempo} A crude lower limit for the relevance of self heating may be obtained by just comparing the film thickness, $d$, with  $\lambda_{th}$. For our samples this leads to \mbox{$t \leq d^2 / 4 D \sim 1$ ns}. Nevertheless, measurements from other groups indicate that other quenching mechanisms, in particular vortex instabilities, could become relevant at much larger characteristic times, of the order of 1 $\mu$s or even larger \cite{Klein1985,Doettinger1994,Sabouret2002,Kunchur2004,Xiao1998,Kunchur2002,LO1976,BS1992,Bernstein2005,Kunchur2000}.
		
\end{thebibliography}
\end{document}